\documentclass[usegraphicx,useAMS,usenatbib]{mn2e}
\usepackage{graphicx,color}
\usepackage{times,gensymb}
\usepackage{amsfonts,amsmath,amssymb,bm,epsfig,mathtools}
\usepackage{verbatim}
\usepackage{multirow}
\usepackage{hyperref}
\usepackage{siunitx}
\usepackage{}

\newcommand{\kms}{\ensuremath{{\rm km\,s}^{-1}}}
\newcommand{\Teff}{\ensuremath{T_{\rm eff}}}
\newcommand{\logg}{\ensuremath{\log(g)}}

\def\newsrc{45}

\begin{document}

\title[Photometric variability of WDs from the PTF]{Photometric variability of candidate white dwarf binary systems from Palomar Transient Factory archival data}
\author[W. Kao et al.]
{Wil~Kao $^{1,2}$\thanks{Present address: Department of Applied Physics, Stanford University, Stanford, CA 94305, USA},
David~L.~Kaplan $^3$,
Thomas~A.~Prince $^1$,
Sumin~Tang $^{1,4}$,
Irina~Ene $^1$,
\newauthor
Kyu Bin~Kim $^1$,
David~Levitan $^1$,
Shrinivas R.~Kulkarni $^1$,
Russ R.~Laher $^5$\\
$^1$ Division of Physics, Mathematics and Astronomy, California Institute of Technology,
Pasadena, CA 91125, USA\\
$^2$ Division of Engineering Science, University of Toronto, 
Toronto, ON, M5S 3H8, Canada\\
$^3$ Department of Physics, University of Wisconsin-Milwaukee, Milwaukee, WI 53211, USA\\
$^4$ Kavli Institute for Theoretical Physics, University of California, 
Santa Barbara, CA 93106, USA\\
$^5$ Spitzer Science Center, California Institute of Technology, M/S 314-6, Pasadena, CA 91125, USA
} 

\maketitle

\begin{abstract}
We present a sample of 59 periodic variables from the Palomar Transient Factory, selected from published catalogues of white dwarf (WD) candidates. 
The variability can likely be attributed to ellipsoidal variation of the tidally distorted companion induced by the gravity of the primary (WD or hot subdwarf) or to the reflection of hot emission by a cooler companion.
We searched \num{11311} spectroscopically or photometrically selected WD candidates from three hot star/WD catalogues, using the Lomb-Scargle periodogram to single out promising sources.
We present period estimates for the candidates, {\newsrc} of which were not previously identified as periodic variables, and find that most have a period shorter than a few days.
Additionally, we discuss the eclipsing systems in our sample and present spectroscopic data on selected sources.
\end{abstract}

\begin{keywords}
	binaries: close - binaries: eclipsing - white dwarfs
\end{keywords}

\section{Introduction}

Most close binary systems containing a white dwarf (WD) star are thought to be a result of one, or possibly two, episodes of unstable mass transfer producing a common envelope that engulfs both the donor star and its companion, called a common envelope (CE) phase \citep{ref:paczynski76, ref:postnov+14}. 
During the CE phase, the core of the donor star and its companion lose angular momentum to the envelope, decreasing the separation of the components of the system.
If sufficient angular momentum and energy can be transferred to the envelope, the envelope can be expelled and spiral-in terminated before merger, resulting in a close post-common envelope binary (PCEB), consisting of the core of the donor star and the companion, typically a main sequence (MS) star.
In fact, hot subdwarf stars of B (sdBs) and O (sdOs) type can assume the role of the donor in such a system as well.
Therefore, depending on the mass of the donor star and its evolution prior to the unstable mass transfer, the PCEB can manifest itself as an MS star together with a WD or sdB/sdO star.
Depending on the type of MS star and the separation of the binary components, the system can further evolve to become a cataclysmic variable (CV), or undergo a second CE phase resulting in a very compact binary comprising two WDs or a WD and a helium star.

A large number of PCEBs have been identified observationally. 
Using the observed PCEB population, comparisons can be made with theoretical estimates of PCEB populations via binary population synthesis codes in order to infer characteristics of the CE phase \citep{ref:davis+10,ref:toonen+nelemans13}.
A useful technique for identifying PCEBs is to test whether a candidate WD is a member of a binary system by either radial velocity measurements, e.g. using the Sloan Digital Sky Survey \citep[SDSS;][]{ref:rebassa-mansergas+12}, or looking for periodic variations in the light curve of the source using large synoptic surveys such as the Palomar Transient Factory \citep[PTF;][]{la} and Catalina Real-Time Transient Sky Survey \citep[CRTS;][]{ma}.

\cite{ref:schreiber+gaensicke03} analysed a sample of 30 PCEBs taken from various literature sources and discussed the age and space density of the PCEB population.
An updated catalogue of WD+MS binaries from SDSS is described in \cite{ref:rebassa-mansergas+12}, and as of 2015 December, the online catalogue lists 203 sources as PCEBs, including 89 with period identifications. 
The web-based catalogue and \cite{ref:rebassa-mansergas+12} provide additional references for discussions of PCEBs found by SDSS in a series of 16 articles between 2007 and 2012.
In particular, a large sample of 58 PCEBs with orbital periods was reported by \citet{go} using SDSS data.
Using CRTS, \citet{par} identified 29 eclipsing WD+MS binaries (12 new) and period estimates were provided for an additional 13 non-eclipsing PCEBs.
More recently, \citet{par2} identified another 17 eclipsing WD+MS binaries (14 new) from SDSS and CRTS using a colour-selected list that targets binaries with cool WDs and/or early M-type MS stars. 
In addition, \citet{dr2} have found several cool WD+M-dwarf binaries with periods less than 0.22~d, and \citet{la2} reported three eclipsing WD+M-dwarf binaries using PTF data.
More generally, both CRTS and PTF have been used to search for PCEBs not limited to WD+MS systems. 
These include eclipsing variables that can be in contact or semidetached\ \citep{dr1} as well as PCEBs with sdB/sdO components rather than WDs \citep{sch}.

In this paper, we present a sample of 59 periodic WD binary candidate systems from the PTF, {\newsrc}  previously unknown.  
For a more complete phase coverage, we cross-checked each object with photometric measurements from the CRTS. 
A period estimate is provided for each of the 59 periodic sources.
The paper is organized as follows.  
Section~\ref{sec:candidates} outlines the WD catalogues that contribute to the list of candidates and provides the selection criteria for the present periodicity search. 
Section~\ref{sec:period} discusses the methods used to detect periodicity as well as the likely astrophysical origins of luminosity variations in compact binary systems.
The eclipsing variables among them are discussed individually.
Section~\ref{sec:spectra} describes spectroscopic measurements on a subset of the periodic sources.
Section~\ref{sec:selection} examines the selection effects and assesses the fidelity of the selection procedure.
Section~\ref{sec:conclusion} provides a summary of the main conclusions of the paper.

\section{Candidate Selection}
\label{sec:candidates}
PTF employs the Palomar $48$-inch Oschin Schmidt Telescope (P48) to search for optical transients and variables and the $60$-inch Telescope (P60) for photometric follow-up. 
The P48 camera has a field of view of $7.26$ $\text{deg}^2$ with a sky coverage of $300$ $\text{deg}^2/\text{hr}$ using cadences from $1$~min to $5$~d, and the 60-s standard exposure time per frame yields a $3\sigma$ limiting magnitude of 20.5 in the $R$ band \citep{la,ra}.
We searched for optical variability in PTF aperture photometry data, processed and calibrated as described in \citet{lah} and further adjusted using relative photometric algorithms from \citet{lev}, for sources in the following, three hot star/WD catalogues -- the spectroscopic WD+MS binary catalogue from the eighth data release (DR8) of SDSS (\citealt{re}), the spectroscopic WD catalogue from SDSS DR7 \citep{kl}, and the catalogue of photometrically selected UV sources and hot WDs from Galaxy Evolution Explorer (GALEX) UV imaging \citep{bi}.
For each source in these catalogues, we searched for matches in the PTF data base with at least 20 measurements in the $R$ band. 
The PTF Mould $R$-band filter is similar in shape to the SDSS $r$-band filter but shifted $27$~\AA\ redward. 
For the present survey, we only considered the $R$-band photometry since most sources have insufficient $g$-band data, but this is likely to change in the future as additional $g$-band observations are carried out.  
To detect strong periodic candidates, we used the Lomb-Scargle (L-S) periodogram \citep{lo,sc}, a least-square spectral analysis method for unevenly sampled data, with modifications for detrending. 
The periodogram retains simple statistical behaviour when the time series is randomly sampled in time. 
We imposed a filtering cut of $3\sigma$ above the median power spectrum density (PSD) on a period search grid between 0.1~d and the sampling time range, and the false positives were ruled out by visually parsing the light curves, as described in Section~\ref{sec:period}.
In addition, we examined those objects with an above-1~mag amplitude of photometric variation in the hope of identifying eclipsing binaries.
Finally, we ruled out artefacts through visual inspection of PTF images.
Table~\ref{table:sources} summarizes the number of selected sources by their catalogue of origin.

In total, the selection procedure yielded 218 variable candidates from the three catalogues, excluding repetitions. 
To improve phase coverage, we supplemented them with CRTS photometry and the six sources not covered by CRTS were excluded.
For the 59 periodic variables in our sample, the median number of measurements is 138 for PTF and 300 for CRTS.
The observation times for both PTF and CRTS data were converted to Modified Heliocentric Julian Date (MHJD) for the subsequent analysis. 

\begin{table*}
	\caption{Variable sources selected from each WD catalogue.
    Note that overlaps do exist among the three catalogues.}
	\centering
	\begin{tabular}{l c c c c}
	\hline
	Catalogue type	&	Total number	&	Number of matches	&	Number of variable sources	&	Reference\\
	\hline
	Spectroscopic WD+MS binaries	 & 	2316	 & 	1316	 &	63	&	\citealt{re}\\
	Spectroscopic WDs	 & 	20407	 & 	2407	 &	41	&	\citealt{kl}\\
	Photometrically selected UV sources	 & 	37347	 & 	10328	 &	174	&	\citealt{bi}\\
	\hline
	\end{tabular}
	\label{table:sources}
\end{table*}

\section{Periodic Candidates}\label{sec:period}
\subsection{Period determination}\label{subsection:period_determination}
Among the 212 variables, we searched for those with ellipsoidal variation and other modes of variability in the light curves.
The L-S periodogram identifies sinusoidal signals, but it can be difficult to associate such signals with a physical mechanism because of unknown harmonic content.
Borrowing the planet-host star discussion by \citet{ja}, one can combine equations from \citet{maz} and \citet{mo} to model the light curve fractional modulation $\Delta F/F$, where $F$ is the mean flux and $\Delta F$ is the modulation amplitude that includes ellipsoidal variation ($A_{\text{ellip}}$), beaming effect ($A_{\text{beam}}$), and reflection ($A_{\text{refl}}$). 
This expression is given by
\begin{equation}\label{eq:folded_lc}
\frac{\Delta F}{F} = -A_{\text{ellip}}\cos(4\pi\phi) + A_{\text{beam}}\sin(2\pi\phi) - A_{\text{refl}}\cos(2\pi\phi),
\end{equation}
where
\begin{equation}\label{eq:parameters}
\begin{aligned}
&A_{\text{ellip}} = \alpha_{\text{ellip}} \frac{m_2 \sin{i}}{m_1} \left(\frac{r_1}{a}\right)^3 \sin{i},\\
&A_{\text{beam}} = 4 \alpha_{\text{beam}} \frac{K_1}{c},\\
&A_{\text{refl}} = 0.1 \alpha_{\text{refl}} \left(\frac{r_2}{a}\right)^2 \sin{i}.
\end{aligned}
\end{equation}
Here $i$ is the orbital inclination, $c$ is the speed of light, $K_1$ and $a$ are the radial velocity semi-amplitude of the primary and binary separation, and $m_1$ ($m_2$) and $r_1$ ($r_2$) are the mass and radius of the primary (companion).
Among the three $\alpha$s, all of order unity, $\alpha_{\text{ellip}}$ depends on the gravity-darkening and limb-darkening coefficients, whereas $\alpha_{\text{beam}}$ corrects the amplitude of the Doppler flux variations for the shifting of flux into and out of the observational passband.
On the other hand, $\alpha_{\text{refl}}$ contains the information about atmospheric properties regarding reflected light and thermal emission.
We define the phase of an observation $\phi_j$ by
\begin{equation}
\phi_j = \frac{t_j-T_0}{P}, 
\end{equation} 
where $t_j$ and $P$ are the epoch and orbital period, respectively.
The lowest magnitude value (highest brightness) in the folded light curve is taken as the reference phase $\phi_0$ and the mean epoch the start time $t_0$.
Each $\phi_0$ corresponds to a physical orientation such that the primary is in inferior conjunction with the companion in the case of reflection effect.
For ellipsoidal modulation, the reference phase coincides with a 90$^{\circ}$ view of the axis connecting the two binary components.
The reference time $T_0$ given in Table~\ref{table:p} is explicitly written as $t_0 + 2\pi\phi_0 P$.

In a typical system, $A_{\text{beam}}$ is small compared to $A_{\text{ellip}}$ and $A_{\text{refl}}$, and for small binary separation, we expect ellipsoidal variation and reflection to be the dominant modes of photometric modulation.
In the case of a double-WD binary, $A_{\text{beam}}$ and $A_{\text{ellip}}$ dominate at periods of a few hours (e.g., NLTT11748; \citealt{shp}) and $\lesssim$ 1~h (e.g., \citealt{br}), respectively.
However, such systems show small photometric amplitudes that are difficult to detect with the typical PTF signal-to-noise ratio (SNR).
According to Eq.~\eqref{eq:parameters}, for ellipsoidal variation, we expect the selection to be biased towards systems with a high companion-to-primary mass ratio and small binary separation. 
The dependence of $\alpha_{\text{ellip}}$ on the gravity-darkening and limb-darkening coefficients means that maximum surface temperature is attained at the poles and minimum at the inner Lagrangian point facing the WD \citep{hy}.
On the other hand, again from Eq.~\eqref{eq:parameters}, compact WD systems are favoured for reflection as well.
However, since the effect stems from the reflection of light from the primary off the companion, big companion size rather than mass is preferred.
The reflection process also implies that the companion is typically cool relative to the more luminous WD primary, whereas ellipsoidal variation can point to a compact high-mass, high-temperature companion such as another WD or a neutron star.

We considered the light curve, L-S periodogram, and folded light curve at the best-fitting L-S period for both PTF and CRTS data.
For the modulation in light due to ellipsoidal variation, we expect to see two minima and two maxima in each folded light curve if we have identified the correct period.
Ellipsoidal variation is mainly a geometric effect; as the companion star goes around the orbit, the observer on Earth sees it face-on (front-back; conjunction) twice and side-on (quadrature) twice per cycle, with more flux seen in the side-on orientation.
Gravity darkening, on the other hand, breaks the symmetry between the two minima as previously alluded to. 
In cases when such symmetry is preserved, the variability is expected to be dominated by reflection, which arises from the difference in brightness between the day side of the companion facing the WD and the night side. 
Unlike ellipsoidal variation, we expect to see one maximum and one minimum in the light curve for each orbital cycle, so there is a factor-of-2 ambiguity in our period determination.
Additional colour and spectral information including GALEX NUV/FUV magnitudes could help guide the choice of modulation mechanism.
We examined all of the fits by eye and picked out those where the depths of the two minima appear significantly different, indicating that the variability is dominated by ellipsoidal variations.  
Given the precision of PTF data, the discrepancy is typically less than $10\%$ of the full magnitude range.
The dominant mode of modulation is highlighted by bold face type in the list of sources given in Table~\ref{table:p}.
However, significant reflection and ellipsoidal variation are unlikely for long-period variables.
Therefore, the dominant mode(s) of photometric modulation for PTF1 J0217--0033, PTF1 J0738+2034, and PTF1 J1359+5538, all of which have a period greater than four days, will be the subject of investigation in a subsequent paper.

Due to scheduling of observations of a given field, the L-S periodogram occasionally picks up ostensibly strong periodicity at periods close to harmonics or subharmonics of a Sidereal day. 
For such a system, we examined the photometry season by season. 
If the seasonal light curve folded at the originally proposed period does not fit well to a sinusoid and a more suitable period was not found, we rejected the candidate. 
A total of eight such sources were removed from the 212 candidates.
In addition, there can be significant power in the L-S periodogram at beat frequencies between the true period and a harmonic or subharmonic of the Sidereal day period. 
Cases where this ambiguity cannot be resolved are indicated in Table~\ref{table:p}.

\begin{figure*}
  \centering
  \includegraphics[width=\linewidth]{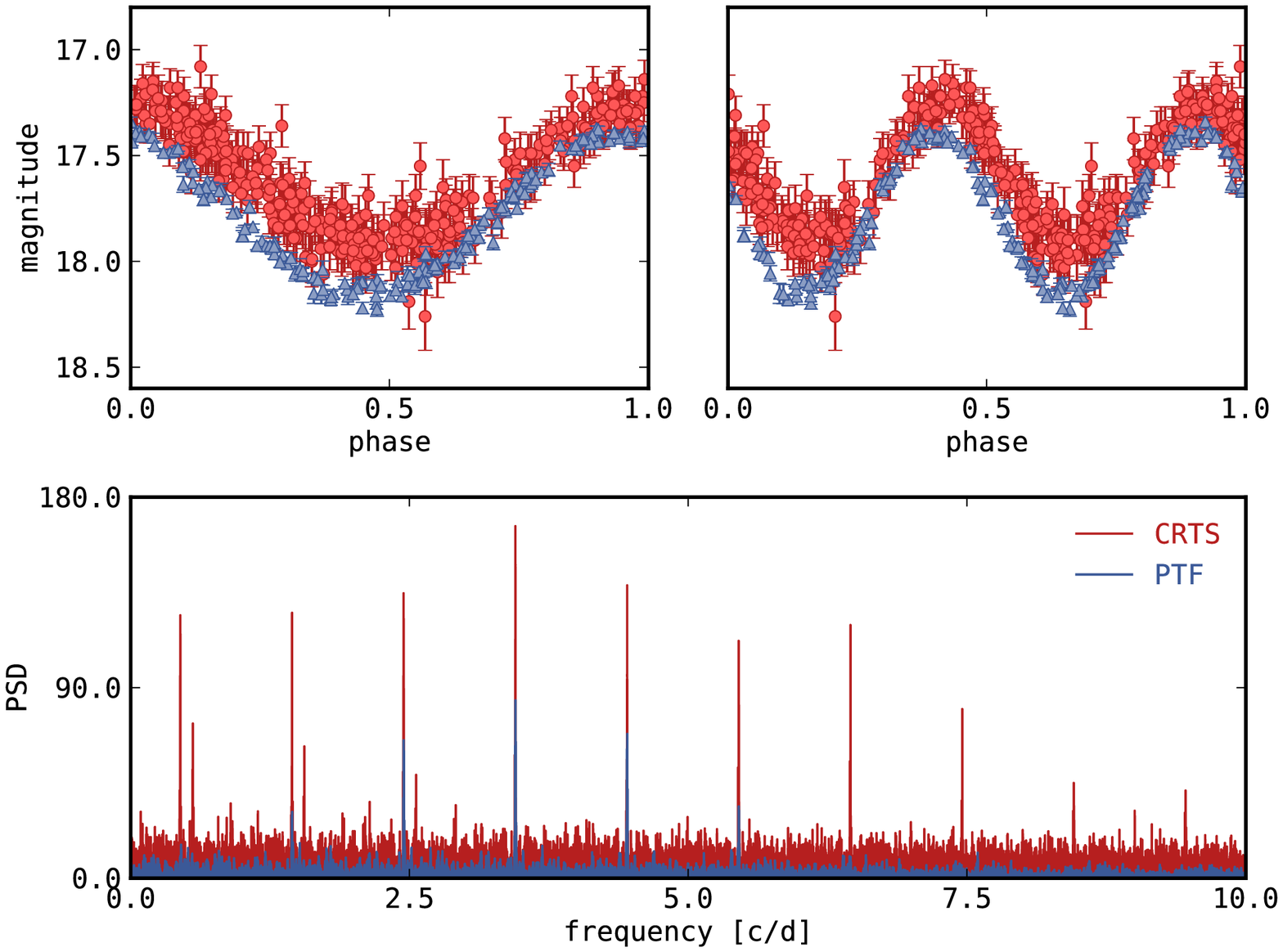}
  \caption{PTF1 J2125-0107. 
  Top-left panel: PTF data folded at $P_{\text{P}}$ and CRTS data folded at $P_{\text{C}}$. Top-right panel: PTF data folded at $2P_{\text{P}}$ and CRTS data folded at $2P_{\text{C}}$.
  Bottom panel: the L-S periodogram.}
  \label{fig:1221aw}
\end{figure*}

A source was accepted if its PTF and CRTS periods differ by $\lesssim 0.1\%$. 
We obtained 23 such systems out of the 204 objects. 
Here we provide an illustrative example in PTF1 J2125--0107. 
Let $P_{\text{P}}$ and $P_{\text{C}}$ be the L-S periods corresponding to the highest L-S PSD for PTF and CRTS respectively.
First, as shown in Fig.~\ref{fig:1221aw}, we have $P_{\text{P}} = P_{\text{C}} = P = 0.28982$~d; clearly, the two periods differ by much less than $0.1\%$.
In the top-left panel, the antinodal positions for both time series very nearly coincide at the same phase.
In addition, the aforementioned gravity darkening effect characteristic of ellipsoidal variation is not pronounced, as shown in the top right panel.
Therefore, we selected $P$ instead of $2P$ as the period estimate, assuming reflection is the dominating mode of modulation.

\begin{figure}
  \centering
  \includegraphics[width=\columnwidth]{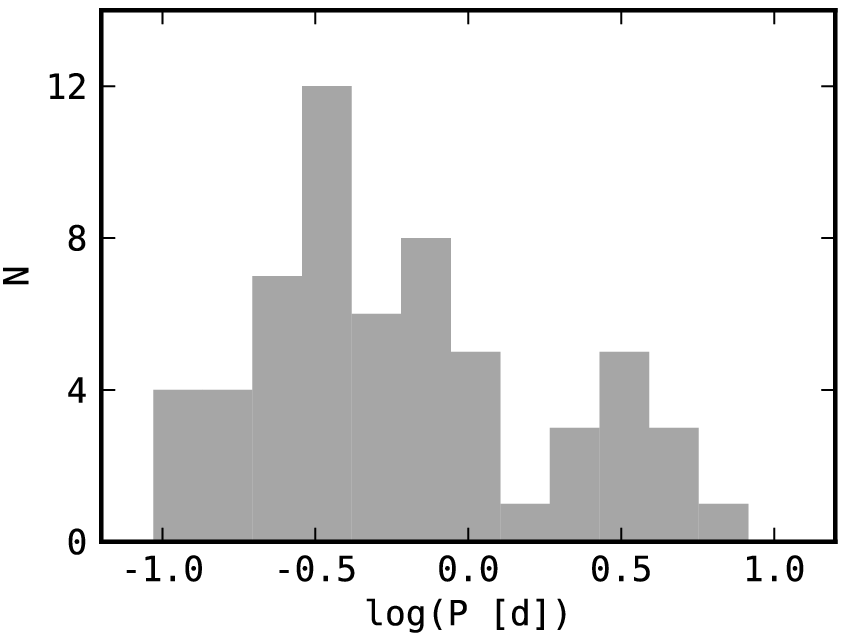}
  \caption{Period distributions of periodic sources listed in Table~\ref{table:p}. }
    \label{fig:p_dis}
\end{figure}

Due to the brighter limiting magnitude typical of CRTS data, we searched the remaining 181 objects using PTF data alone.
There are 36 additional objects that were found to exhibit significant periodic modulation.
To determine the true period, we visually evaluated the light curves folded at the top five L-S periods by considering the degree of scattering within each phase bin.
All 59 light curves folded by the chosen period can be found in Appendix~\ref{section:flc} and the relevant parameters are listed in Table~\ref{table:p}.
The error in each period estimate was determined by first computing the $\chi^2$ statistics of the L-S fit and constructing $\chi^2$ as a function of frequency $f$.
We then fit a parabola to $\chi^2(f)$ near the minimum $\chi^2$ point in order to interpolate for the best-fitting uncertainty.
The resulting period distribution is shown in Fig.~\ref{fig:p_dis}.
The mode of the distribution lies in the compact end (i.e., $P \lesssim 1$~d) with a tail extended to around 10 days.

\begin{table*}
	\centering
	\caption{Periodic variables: $P_\text{e}$ and $P_\text{r}$ are the periods from ellipsoidal modulation and reflection, respectively.
	The reference time $T_0$ is taken as the sum of the mean time span and time converted from the reference phase corresponding to the photometric maximum in the folded light curve.
	}
	\begin{tabular}{l c c c c c c c c c}
	\hline
Name$^{\text{(a)}}$	&	$P_\text{e}$ (d)	&	$P_\text{r}$ (d)	&	$T_0$ (MHJD)	&	$R$ (mag)	&	$\Delta R$ (mag)	&	$T_{\text{eff}}^{\text{(b)}}$ (K)	&	$\log{g}^{\text{(b)}}$	&	Class.	&	Ctlg.$^{\text{(c)}}$\\
\hline																		
PTF1 J000152.09+000644.3	&	1.331424(9)	&	\textbf{0.665712(9)}	&	55051.0(2)	&	17.83	&	0.1	&	40000$^\text{m}$	&	7.5$^\text{m}$	&	WD+K	&	B\\
PTF1 J011339.09+225739.1$^\text{e}$	&	0.18674600(5)	&	\textbf{0.09337300(5)}	&	55481.7(5)	&	16.98	&	0.56	&	29980	&	5.69	&	sdB+M 	&	B\\
PTF1 J021726.32--003317.8	&	\textbf{5.3703(2)}	&	2.6852(2)	&	55063.1(2)	&	16.12	&	0.05	&	6300$^\text{k}$	&	5.00$^\text{k}$	&	DA+M	&	KR\\
PTF1 J022349.45+215946.2	&	1.3642(2)	&	\textbf{0.6821(2)}	&	55484.45(4)	&	15.06	&	0.1	&	--	&	--	&	--	&	B\\
PTF1 J025403.75+005854.2	&	2.174410(5)	&	\textbf{1.087205(5)}	&	55057.53(1)	&	17.9	&	0.32	&	100000$^\text{k}$	&	7.21$^\text{k}$	&	--	&	BDK\\
PTF1 J031452.10+020607.1$^{\text{ce}}$	&	0.610590(1)	&	\textbf{0.305295(1)}	&	55055.4(5)	&	17.33	&	0.25	&	41140$^\text{l}$	&	8.02$^\text{l}$	&	--	&	B\\
PTF1 J073853.58+203446.2	&	\num{9.2389(3)}	&	\textbf{4.6194(3)}	&	55815.7(2)	&	15.8	&	0.11	&	--	&	--	&	--	&	B\\
PTF1 J074111.48+215554.6	&	0.777373(2)	&	\textbf{0.388686(2)}	&	55813.47(2)	&	15.96	&	0.17	&	--	&	--	&	--	&	B\\
PTF1 J080940.38+453357.0	&	\textbf{0.28378231(6)}	&	0.14189116(6)	&	55080.98(2)	&	14.07	&	0.13	&	--	&	--	&	--	&	BD\\
PTF1 J081606.68+455525.5	&	0.626479(3)	&	\textbf{0.313240(3)}	&	54907.72(2)	&	16.17	&	0.05	&	--	&	--	&	--	&	B\\
PTF1 J082005.22+210432.5	&	0.5740531(6)	&	\textbf{0.2870266(6)}	&	54907.14(1)	&	16.13	&	0.08	&	--	&	--	&	--	&	BD\\
PTF1 J082823.58+210036.0	&	0.683363(4)	&	\textbf{0.341681(4)}	&	54907.14(6)	&	19.26	&	0.32	&	28718$^\text{r}$	&	9.13$^\text{r}$	&	DA+M2	&	BR\\
PTF1 J084426.84+221155.7	&	1.553(2)	&	\textbf{0.776(2)}	&	55314.53(5)	&	16.68	&	0.09	&	--	&	--	&	--	&	B\\
PTF1 J085137.18+290330.2	&	1.008499(5)	&	\textbf{0.504250(5)}	&	54907.03(9)	&	15	&	0.09	&	--	&	--	&	--	&	B\\
PTF1 J085414.26+211148.2	&	\textbf{0.204307(6)}	&	0.102153(6)	&	55903.45(3)	&	15.18	&	0.11	&	--	&	--	&	--	&	BDP\\
PTF1 J095306.83+013817.7	&	6.5800(2)	&	\textbf{3.2900(2)}	&	54911.64(5)	&	17.11	&	0.16	&	--	&	--	&	--	&	B\\
PTF1 J102113.90+471003.5	&	0.6286995(8)	&	\textbf{0.3143497(8)}	&	54907.2(1)	&	16.72	&	0.12	&	--	&	--	&	--	&	B\\
PTF1 J103258.79+332529.9	&	1.3097(1)	&	\textbf{0.6549(1)}	&	56308.0(1)	&	15.68	&	0.1	&	--	&	--	&	--	&	B\\
PTF1 J114509.77+381329.3	&	\textbf{0.19003580(5)}	&	0.09501790(5)	&	54906.1(5)	&	15.67	&	0.12	&	7000$^\text{m}$	&	8.25$^\text{m}$	&	DA+M	&	PR\\
PTF1 J115744.84+485618.2$^{\text{ce}}$	&	4.582356(7)	&	\textbf{2.291178(7)}	&	55976.3(1)	&	15.33	&	1	&	--	&	--	&	--	&	BD\\
PTF1 J122930.65+263050.5	&	0.671151(7)	&	\textbf{0.335576(7)}	&	54973.47(3)	&	17.09	&	0.06	&	21289$^\text{r}$	&	8.64$^\text{r}$	&	DA+M	&	KPR\\
PTF1 J123159.53+670918.9$^{\text{pe}}$	&	0.2259981(4)	&	\textbf{0.1129991(4)}	&	54959.4(2)	&	18.46	&	0.57	&	35740$^\text{r}$	&	7.38$^\text{r}$	&	DA+M	&	BKR\\
PTF1 J123309.63+083434.5	&	5.5864(1)	&	\textbf{2.7932(1)}	&	54939.6(2)	&	15.53	&	0.08	&	--	&	--	&	--	&	B\\
PTF1 J123339.39+135943.8	&	0.811004(3)	&	\textbf{0.405502(3)}	&	54980.7(4)	&	17.1	&	0.11	&	--	&	--	&	--	&	B\\
PTF1 J130733.50+215636.6$^\text{e}$	&	\textbf{0.2163225(4)}	&	0.1081612(4)	&	54908.6(1)	&	17.19	&	0.12	&	--	&	--	&	DC+M4	&	PR\\
PTF1 J131751.64+673159.2	&	6.7642(3)	&	\textbf{3.3821(3)}	&	54963.0(3)	&	16.13	&	0.05	&	99575$^\text{r}$	&	8.18$^\text{r}$	&	DA+M	&	KR\\
PTF1 J134240.40+293430.1	&	1.502847(2)	&	\textbf{0.751423(2)}	&	54905.18(7)	&	16.18	&	0.12	&	--	&	--	&	--	&	B\\
PTF1 J135016.01+602437.7	&	1.307684(6)	&	\textbf{0.653842(6)}	&	54908.91(3)	&	16.7	&	0.09	&	--	&	--	&	--	&	B\\
PTF1 J135922.51+553836.7	&	16.466(1)	&	\textbf{8.233(1)}	&	54931.59(5)	&	16.69	&	0.13	&	--	&	--	&	--	&	B\\
PTF1 J141602.87+372806.8	&	2.22962(2)	&	\textbf{1.11481(2)}	&	54906.92(5)	&	17.09	&	0.08	&	--	&	--	&	--	&	B\\
PTF1 J150525.34+070635.6	&	2.289647(7)	&	\textbf{1.144823(7)}	&	55012.23(9)	&	18.09	&	0.18	&	--	&	--	&	--	&	B\\
PTF1 J151227.81+013934.5	&	0.85800(1)	&	\textbf{0.42900(1)}	&	54907.922(8)	&	17.22	&	0.16	&	--	&	--	&	--	&	B\\
PTF1 J151500.57+191619.8	&	0.242951(3)	&	\textbf{0.121476(3)}	&	56078.2(4)	&	18.04	&	0.47	&	30000$^\text{g}$	&	8.00$^\text{g}$	&	--	&	BD\\
PTF1 J151706.31+053035.5	&	1.399109(6)	&	\textbf{0.699554(6)}	&	55012.5(1)	&	15.01	&	0.06	&	--	&	--	&	--	&	B\\
PTF1 J152416.95+504748.8	&	0.4814238(5)	&	\textbf{0.2407119(5)}	&	54958.21(1)	&	16.28	&	0.08	&	18000$^\text{m}$	&	8.25$^\text{m}$	&	DA+M2	&	R\\
PTF1 J153938.10+270605.8	&	0.477086(4)	&	\textbf{0.238543(4)}	&	55403.7(9)	&	16.94	&	0.07	&	36572$^\text{r}$	&	7.31$^\text{r}$	&	DA+M	&	KR\\
PTF1 J154434.95+095451.4	&	5.5530(1)	&	\textbf{2.7765(1)}	&	55005.4(7)	&	16.44	&	0.1	&	--	&	--	&	--	&	B\\
PTF1 J155256.11+125443.9$^{\text{ce}}$	&	\textbf{0.2601610(1)}	&	0.1300805(1)	&	54980.73(5)	&	16.78	&	0.48	&	--	&	--	&	--	&	B\\
PTF1 J155904.62+035623.5$^\text{a}$	&	0.188695(5)	&	\textbf{0.094348(5)}	&	56077.9(3)	&	18.8	&	0.32	&	48770$^\text{r}$	&	7.98$^\text{r}$	&	DA+M	&	BKPR\\
PTF1 J160540.13+461046.0	&	\textbf{4.7659(1)}	&	2.3830(1)	&	54970.1(3)	&	16.48	&	0.06	&	31853$^\text{r}$	&	8.03$^\text{r}$	&	DA+M	&	KR\\
PTF1 J161129.25+280626.3	&	\textbf{0.4990411(9)}	&	0.2495205(9)	&	54972.1(1)	&	16.74	&	0.29	&	--	&	--	&	--	&	B\\
PTF1 J162028.94+630446.7	&	0.598858(5)	&	\textbf{0.299429(5)}	&	54960.5(9)	&	18.73	&	0.18	&	23551$^\text{k}$	&	7.12$^\text{k}$	&	DA+M	&	KR\\
PTF1 J162035.14+421542.2	&	4.59325(7)	&	\textbf{2.29662(7)}	&	54965.45(9)	&	16.42	&	0.06	&	--	&	--	&	--	&	B\\
PTF1 J162209.32+500752.5	&	0.646663(1)	&	\textbf{0.323331(1)}	&	54959.54(1)	&	16.94	&	0.09	&	30000$^\text{m}$	&	8.0$^\text{m}$	&	WD+M1	&	BR\\
PTF1 J162351.64+403211.3	&	1.181591(7)	&	\textbf{0.590795(7)}	&	54960.91(8)	&	18.32	&	0.18	&	48827$^\text{s}$	&	7.9$^\text{s}$	&	DA+M	&	BKR\\
PTF1 J162821.79+315726.0	&	0.4578609(6)	&	\textbf{0.2289304(6)}	&	54964.313(7)	&	15.91	&	0.08	&	--	&	--	&	WD+K	&	B\\
PTF1 J164519.45+445736.3	&	0.710548(1)	&	\textbf{0.355274(1)}	&	54957.3(1)	&	16.99	&	0.12	&	--	&	--	&	--	&	B\\
PTF1 J172406.14+562003.1	&	\textbf{0.666039(2)}	&	0.333019(2)	&	55349.36(5)	&	16.65	&	0.12	&	36250$^\text{s}$	&	7.2$^\text{s}$	&	DA+M	&	RK\\
PTF1 J173002.48+333401.9	&	0.3138968(4)	&	\textbf{0.1569484(4)}	&	54964.915(9)	&	18.61	&	0.3	&	47114$^\text{r}$	&	7.35$^\text{r}$	&	DA+M	&	BDPR\\
PTF1 J173338.15+564432.4	&	4.1973(1)	&	\textbf{2.0987(1)}	&	55355.43(7)	&	14.87	&	0.07	&	--	&	--	&	--	&	B\\
PTF1 J204909.19+002604.2	&	2.13223(1)	&	\textbf{1.06611(1)}	&	55008.85(3)	&	17.14	&	0.09	&	--	&	--	&	WD+K	&	B\\
PTF1 J212531.92--010745.8	&	0.5796404(2)	&	\textbf{0.2898202(2)}	&	55008.65(1)	&	17.85	&	0.44	&	12692$^\text{k}$	&	9.98$^\text{k}$	&	DAB+M	&	BDKR\\
PTF1 J213941.46+002747.2	&	3.23341(6)	&	\textbf{1.61670(6)}	&	55011.9(5)	&	17.2	&	0.09	&	--	&	--	&	WD+K	&	B\\
PTF1 J221804.58+415149.3	&	0.3742598(6)	&	\textbf{0.1871299(6)}	&	55006.985(2)	&	19.51	&	0.51	&	--	&	--	&	--	&	B\\
PTF1 J223114.66+344125.6	&	1.145789(7)	&	\textbf{0.572895(7)}	&	55009.767(9)	&	16.56	&	0.1	&	--	&	--	&	---	&	B\\
PTF1 J223530.61+142855.0$^\text{e}$	&	0.288915(2)	&	\textbf{0.144457(2)}	&	55019.95(3)	&	18.9	&	0.18	&	21277$^\text{s}$	&	7.6$^\text{s}$	&	DA+M	&	BKPR\\
PTF1 J225256.21--000406.0	&	2.02167(1)	&	\textbf{1.01083(1)}	&	55011.43(3)	&	17.17	&	0.12	&	--	&	--	&	WD+K	&	B\\
PTF1 J231254.41--000129.0	&	6.0566(1)	&	\textbf{3.0283(1)}	&	55403.8(2)	&	14.07	&	0.28	&	--	&	--	&	WD+K	&	B\\
PTF1 J232730.79+070115.1	&	\textbf{0.5664410(8)}	&	0.2832205(8)	&	55010.1(2)	&	18.62	&	0.32	&	--	&	--	&	--	&	B\\
&	&	&	&	&	&	&	&	&\\
\multicolumn{10}{p{18cm}}{
(a) e -- Eclipses confirmed by \citet{par} and/or PTF images; pe/ce -- Eclipses identified solely by PTF/CRTS photometry; a -- Unresolved aliases in period.
(b) g -- \citet{gi}; k -- \citet{kl}; l -- \citet{li};  m -- \citet{mor}; r -- \citet{re}; s --  \citet{si}.
(c) B -- \citet{bi}; D -- \citet{dr1,dr2}; K -- \citet{kl}; P -- \citet{par,par2}; and R -- \citet{re}.}\\
	\hline
	\end{tabular}
	\label{table:p}
\end{table*}

\subsection{Previously identified sources}\label{subsection:known}
We searched other catalogues for matches with sources that have been identified as periodic variables. 
In particular, we obtained eight matches with \citet{dr1}. 
\cite{par,par2} previously identified three of the eclipsing sources in Table~\ref{table:ref_p} (PTF1 J0745+2631, PTF1 J1307+2156 and PTF1 J2235+1428).
\citet{par} also reported a catalogue of non-eclipsing PCEBs from the Catalina Sky Survey photometry, and four were identified by our search (PTF1 J1145+3813, PTF1 J1229+2630, PTF1 J1559+0356, and PTF1 J1730+3334).
Furthermore, we note that PTF1 J0125--0107 has been determined by \citet{na} as the first definite close binary system containing a pre-degenerate (PG 1159) star.
These known periodic variables are summarized in Table~\ref{table:known}.
After resolving the doubling ambiguity, we found that all period determinations in Table~\ref{table:known} agree with previous works to at least one part per thousand, and in fact all but two have been determined more accurately than in previous works.

Additionally, we cross-checked periodic variables in our sample against sources in SIMBAD with a search radius of 3~arcsec. 
The search yielded 27 matches which are listed in Table~\ref{table:ref_p}.
The SIMBAD identifiers may be helpful for finding additional literature references on the cross-matched sources.

\subsection{Eclipsing systems}\label{subsection:eclipsing_systems}
\begin{figure}
 \centering
 \includegraphics[width=\columnwidth]{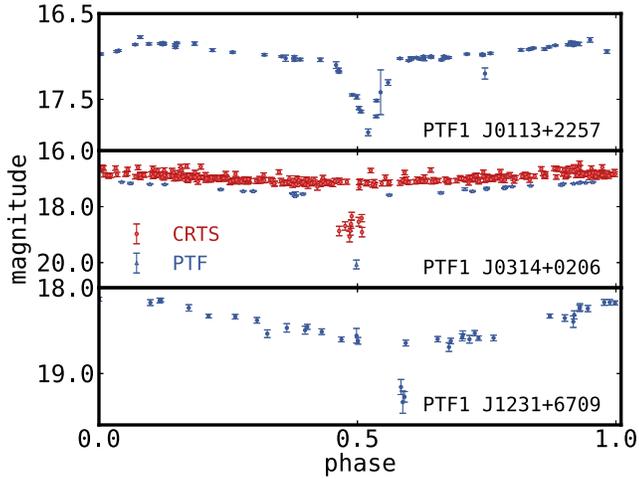}
 \caption{Light curves of the three new eclipsing systems, folded by the L-S period given in Table~\ref{table:p}.}
 \label{fig:e}
\end{figure}

Among the 59 periodic candidates, we identified six eclipsing systems, three of which were not previously reported.
These new eclipsing binaries are shown in Fig.~\ref{fig:e}.
 
\subsubsection{PTF1 J011339.09+225739.1}
PTF1 J0113+2257 is a new eclipsing source identified by PTF photometry.
From the L-S periodogram, we obtained $P = 0.09337300(5)$~d, the shortest period of any source in our sample.
As suggested by the short period as well as spectroscopy (see Section~\ref{sec:spectra}), this source is likely an sdB+M system. 

\subsubsection{PTF1 J031452.10+020607.1}
A single eclipse point is observed in PTF and multiple eclipse points in CRTS data.
The earliest and latest points in the eclipse observed in CRTS are at phase 0.46 and 0.51, respectively, leading to a lower bound of $\approx 0.3$~h for the eclipse duration.

\subsubsection{PTF1J115744.84+485618.2}
PTF1 J1157+4856 is an eclipsing binary, designated as BE UMa. 
It is a PCEB consisting of a hot subdwarf and a cool G--K dwarf with a period of $2.29$~d \citep{sh}. 
The photometric variability is dominated by reflection \citep{mar}.

\subsubsection{PTF1 J123159.53+670918.9}
PTF1 J1231+6709 is another new eclipsing source identified by PTF photometry.
It was originally designated as an SDSS PCEB \citep{re}.
From the L-S periodogram, we obtained $P = 0.1129991(4)$~d, and the depth of the eclipse is approximately $1$~mag.

\subsubsection{PTF1 J130733.50+215636.6}
PTF1 J1307+2156 was identified by \citet{par} as an eclipsing PCEB with a period of 0.2163221322(1)~d. 
The eclipse was captured in both PTF and CRTS data, and L-S periodogram gives $P = 0.2163225(4)$~d, which confirms the period estimate reported by \citet{par}. 

\subsubsection{PTF1 J223530.61+142855.0}
\citet{par} reported a period of 0.1444564852(34)~d for PTF1 J2235+1428 which is consistent with our L-S best-fitting period of 0.144457(2)~d.
In Section~\ref{sec:spectra}, we provide refined WD parameters via spectroscopic analysis.

\begin{table*}
	\caption{Comparison between the periods reported in this work and the literature.
	}
	\begin{tabular}{ l  c  c }
	\hline
	Name &	This work (d)  &	Lit. (d) \\
	\hline
	PTF1 0254+0058 (D)	&	1.087205(5) 	&	2.1744226\\
	PTF1 0809+4533 (D)	&	0.28378231(6)	&	0.283774 \\
	PTF1 0820+2104 (D)	&	0.2870266(6)	&	0.2870273\\
	PTF1 0854+2111 (DP)		&	0.204307(6)	&	0.20430622\\
	PTF1 1145+3813 (P)		&	0.19003580(5)	&	0.19003799(27)\\	
	PTF1 1157+4856 (D)	&	2.291178(7)	&	2.2909892\\	
	PTF1 1229+2630 (P)	&	0.335576(7)	&	0.671148(66)\\
	PTF1 1307+2156 (P)	&	0.2163225(4)	&	0.2163221322(1)\\
	PTF1 1515+1916 (D)	&	0.121476(3)	&	0.121435\\
	PTF1 1559+0356 (P)	&	0.094348(5)	&	0.0943473(1)\\
	PTF1 1724+5620 (R)	&	0.666039(2)	&	0.3330193(1)\\
	PTF1 1730+3334 (DP)	&	0.1569484(4)	&	0.1569473(3)\\
	PTF1 2125--0107 (DN)		&	0.2898202(2)	&	0.289817\\
	PTF1 2235+1428 (P)	&	0.144457(2)	&	0.1444564852(34)\\
	&	&	\\
\multicolumn{3}{p{10cm}}{D - \citet{dr1,dr2};  N - \citet{na}; P - \citet{par,par2}; R - \citet{ref:rebassa-mansergas+12}. 
	The most updated period from the literature is presented in the case of a conflict.}\\
	\hline
	\end{tabular}
	\label{table:known}
\end{table*}

\begin{table*}
	\centering
	\caption{Alternative source names from SIMBAD.}
	\begin{tabular}{l c c c}
	\hline
	Name	&	SIMBAD identifier	&	Dist (arcsec)	&	Spectral type\\
	\hline
	PTF1 J000152.09+000644.3	&	2MASS J00015207+0006445	&	0.37	&		dM0+DA C\\
	PTF1 J011339.09+225739.1	&	SDSS J011339.09+225739.0	&	0.09	&		\\
	PTF1 J021726.32-003317.8	&	2MASS J02172631-0033179	&	0.18	&		DA+M\\
	PTF1 J025403.75+005854.2	&	SDSS J025403.75+005854.4	&	0.04	&		DO D\\
	PTF1 J031452.10+020607.1	&	WD 0312+019	&	0.38	&		DA1.2 C\\
	PTF1 J082823.58+210036.0	&	SDSS J082823.58+210036.0	&	0.05	&		M2 D\\
	PTF1 J085414.26+211148.2	&	2MASS J08541431+2111483	&	0.7	&		\\
	PTF1 J114509.77+381329.3	&	2MASS J11450979+3813292	&	0.28	&		DA+M C\\
	PTF1 J115744.84+485618.2	&	V* BE UMa	&	0.32	&		CV D\\
	PTF1 J122930.65+263050.5	&	2MASS J12293066+2630503	&	0.18	&		DA+M D\\
	PTF1 J123159.53+670918.9	&	WD 1229+674	&	0.16	&		DA D\\
	PTF1 J123339.39+135943.8	&	2MASS J12333939+1359439	&	0.13	&		\\	
	PTF1 J130733.50+215636.6	&	2MASS J13073350+2156370	&	0.5	&		DC+M4 D\\
	PTF1 J131751.64+673159.2	&	[HHD2009] SDSS J1317+6731 WD	&	0.41	&		DA\\
	PTF1 J141602.87+372806.8	&	2MASS J14160286+3728069	&	0.24	&		\\
	PTF1 J151500.57+191619.8	&	SDSS J151500.57+191619.6	&	0.08	&	DA+dM\\
	PTF1 J152416.95+504748.8	&	2MASS J15241697+5047489	&	0.25	&		DA+M2 D\\
	PTF1 J153938.10+270605.8	&	Ton 243	&	0.05	&		DA+M D\\
	PTF1 J155256.11+125443.9	&	V* NN Ser	&	0.9	&		DAO1+M4\\
	PTF1 J155904.62+035623.5	&	SDSS J155904.62+035623.4	&	0.07	&		DA+M D\\
	PTF1 J160540.13+461046.0	&	[HHD2009] SDSS J1605+4610 WD	&	0.39	&		DA\\					
    PTF1 J162209.32+500752.5	&	2MASS J16220932+5007524	&	0.09	&		WD+M1 D\\
	PTF1 J162351.64+403211.3	&	SDSS J162351.64+403211.3	&	0.05	&		DA+dM D\\
	PTF1 J172406.14+562003.1	&	2MASS J17240613+5620033	&	0.03	&		DA+:e\\
	PTF1 J173002.48+333401.9	&	SDSS J173002.48+333401.8	&	0.06	&		\\
	PTF1 J212531.92-010745.8	&	[HHD2009] SDSS J2125-0107 WD	&	0.11	&		DO\\
	PTF1 J223530.61+142855.0	&	SDSS J223530.61+142855.1	&	0.2	&		DA+dM:e D\\
	\hline
	\end{tabular}
	\label{table:ref_p}
\end{table*}


\section{Spectroscopy of Selected Systems}\label{sec:spectra}

We obtained multiple epochs of low-resolution ($R \sim 1500$) spectroscopy to determine or refine the WD atmospheric parameters including the effective temperature, surface gravity, and possible composition for selected sources.
More importantly, these spectra helped confirm through radial velocity variation that the observed photometric variability originated from orbital motion.
Here we discuss spectroscopic studies of three selected objects that highlight the different classes of sources in our sample. 
All sources were observed with the Double Spectrograph (DBSP) on the 200-inch Hale Telescope (P200) at the Palomar Observatory.  
A summary of target observations is provided in Table~\ref{table:obs}.
Our standard observing configuration used a dichroic at 5500\,\AA\ to separate the light into blue and red channels and a 600 lines/mm grating on the blue side which covers roughly 3800--5500\,\AA.
We concentrate on the blue channel, since given the hot nature and hydrogen atmospheres of most targets, only H $\alpha$ is visible on the red channel and the overall SNR is lower.

We used a grid of WD model atmospheres based on \citet{koester10}, with the modification of using ML2/$\alpha=0.8$ for parameterizing the convective energy transport instead of ML2/$\alpha=0.6$.  
These models cover hydrogen-atmosphere (DA) white dwarfs with surface gravities $\log(g)$ between 7.0 and 10.0 in steps of 0.25\,dex and effective temperatures between 6000\,K and \num{60000}\,K in steps of 250\,K (up to \num{20000}\,K), 1000\,K (up to \num{30000}\,K), and 5000\,K.  
We have additional models \citep{koester10} for sdBs with $\log(g)$ of 5.0 and 6.0, surface temperatures of \num{14000}\,K--\num{26000}\,K in steps of 1000\,K, and surface helium abundances of $\log {\rm He/H}=0$, $-0.3$, and $-1.0$.
We used a \textsc{PyRAF}-based spectral reduction pipeline for DBSP \footnote{http://github.com/ebellm/pyraf-dbsp} to perform debiasing, flat fielding, and wavelength and flux calibration.

\begin{table*}
	\centering
	\caption{Summary of target spectroscopic observations.}
	\begin{tabular}{l c c c c}
	\hline
	Name	&	\# Obs.	&	Exposure (s)	&	UT date	& Instrument\\
	\hline
	PTF1 J223530.61+142855.0	&	4	&	900	&	2013 July 15	&	P200+DBSP\\
	PTF1 J173002.48+333401.9	&	5	&	900	&	2013 July 15	&	P200+DBSP\\
	PTF1 J011339.09+225739.1	&	1	&	300	&	2014 July 30	&	P200+DBSP\\
							&	3	&	600	&	2014 October 31	&	P200+DBSP\\
	\hline
	\end{tabular}
	\label{table:obs}
\end{table*}

\begin{figure*}
 \centering
 \includegraphics[width=\textwidth]{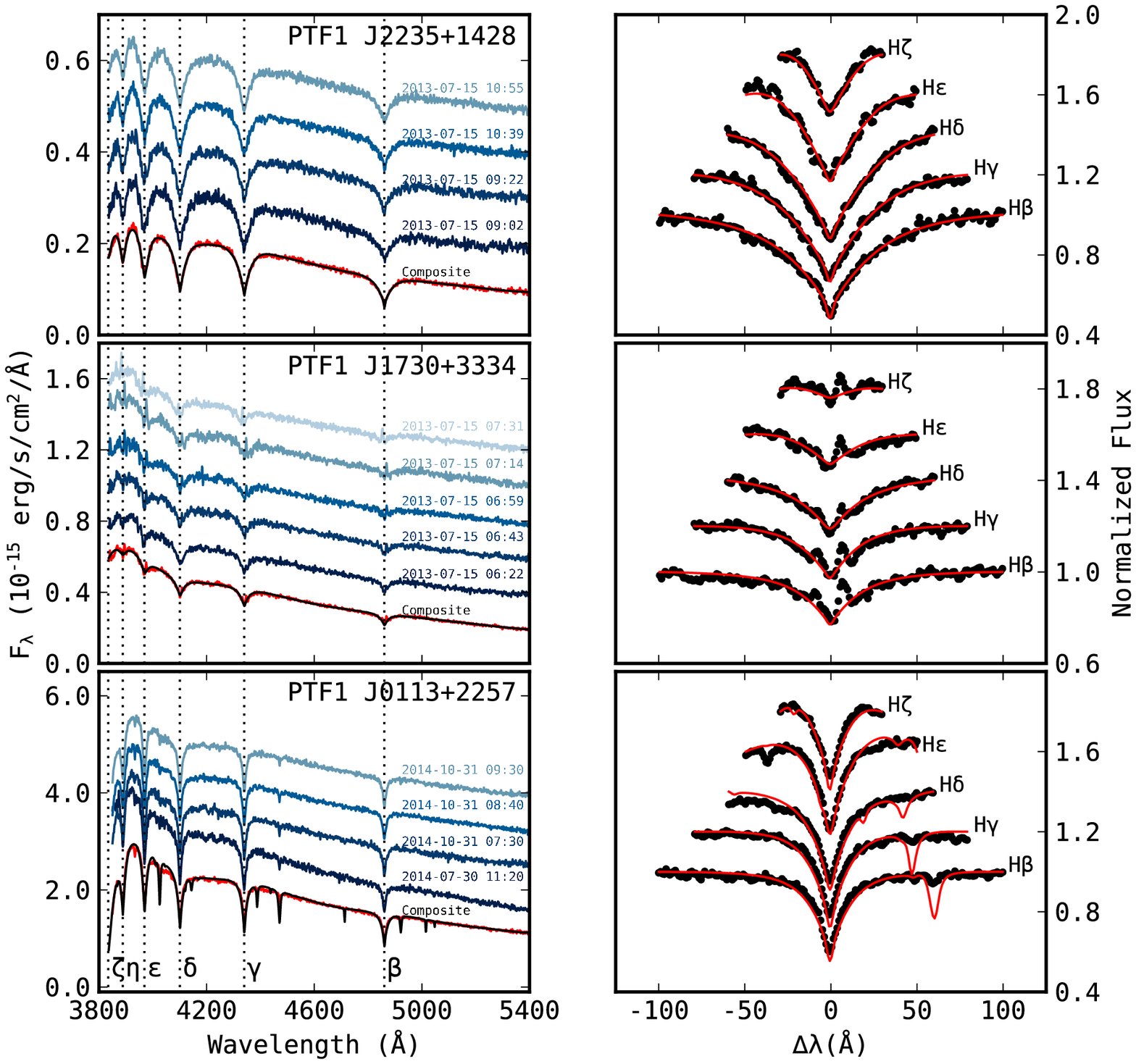}
 \caption{Low-resolution spectroscopy for selected sources. 
Left-hand panels: DBSP spectra (labelled by UT), all shifted to zero radial velocity and vertically offset, alongside our composite spectrum with the best-fit model (see the text) overplotted.
The wavelengths of the Balmer series are shown by the vertical lines.
Right-hand panels: fits to the Balmer lines with the composite spectrum overplotted.
}
 \label{fig:spectra}
\end{figure*}

\subsection{PTF1 J173002.48+333401.9, a low-mass WD with large RV amplitude}
This source was observed five times with an exposure of 900\,s each on the night of 2013-July-15.  
The observations were consecutive exposures covering roughly 84\,min including readout, so we covered roughly 40\% of the 0.16\,d orbit.  
As seen in Fig.~\ref{fig:spectra}, the spectra resemble that of a hot WD with some Balmer emission lines superimposed.  
First, we fitted for the radial velocity of the Balmer emission lines using narrow Gaussians as the template and found best-fitting velocities for each spectrum from $+350$ to $-36\,\kms$.  
We then masked the portion of each spectrum affected by the Balmer emission lines and used the remaining data to fit a hot WD atmosphere.  
We obtained $\Teff=\num{45000}$\,K and $\logg=7.5$, our closest model to the published values of $\logg = 7.35$ and $\Teff = \num{47114}$\,K \citep{re}.    
Each model was convolved with the spectral response of our data, described by a Gaussian with a full width at half-maximum of 5\,\AA\ set by the typical seeing of 1.8 arcsec and truncated at 1.5 arcsec to mimic the slit (see e.g. \citealt{ka}).  
We then used an 11th order polynomial to fit the continuum.  
The fits are good, with typical $\chi^2=1662$ for 1670 degrees of freedom. 
The resulting velocities range from $-144$ to $+40\,\kms$, with typical uncertainties of $\pm25\,\kms$.  
It is evident from Fig.~\ref{fig:spectra} that the velocities of the Balmer emission lines are not the same as the WD atmosphere fits, suggesting that they originated in different parts of the binary system.  
In fact, the emission-line velocities are roughly $180\degr$ out of phase with the atmosphere fits.  
A likely explanation is that the emission lines can be attributed to the emission from the heated day side of a low-mass companion.  
We estimate the WD-to-companion mass ratio to be roughly 2 to 3 from the radial velocity amplitude ratio of the Balmer emission lines compared to the WD absorption lines, although given the poor phase coverage, we cannot be more precise.
The surface gravity and temperature imply a mass of about 0.45\,$M\odot$ for the WD when compared to the DA evolutionary models from \citealt{tbg11}, so the companion mass is around 0.2\,$M\odot$.

\subsection{PTF1 J011339.09+225739.1 and PTF1 J223530.61+142855.0, eclipsing binaries}
PTF1 J2235+1428, observed four times with an exposure of 900\,s each on the night of 2013-July-15, was reported by \citet{par} as a deep ($\sim 3$ mag) eclipsing PCEB.  
We had two pairs of consecutive exposures separated by roughly 75\,min, so the total span of the observations is slightly more than half of an orbit.
Similar to PTF1 J1730+3334, the spectra resemble that of a hot WD with some Balmer emission lines superimposed, as shown in Fig.~\ref{fig:spectra}.
We fit the data using a WD atmosphere model with $\Teff=\num{21000}$\,K and $\logg=7.5$, our closest model to the published values of $\logg = 7.6$ and $\Teff = \num{21277}$\,K \citep{si}, using an identical fitting procedure as PTF1 J1730+3334.
The fits are good, with typical $\chi^2=1196$ for 1670 degrees of freedom. 
We found that the first pair of exposures have similar velocities of $-61\pm12$ and $-81\pm12\,\kms$, while the second pair are significant shifted ($138\pm12$ and $-55\pm12\,\kms$).  

PTF1 J0113+2257 is a newly discovered eclipsing binary that has the shortest orbital period ($P = 0.09337300(5)$~d) among the 59 periodic candidates in our catalogue.  
This source was observed four times---once on 2014 July 30 for 300\,s and three times on 2014 October 31 for 600\,s each, separated by roughly one hour.  
It appears to be a subdwarf with visible lines of He~\textsc{I} in addition to the broad Balmer lines.  
Using a series of sdB atmospheres covering $\Teff=\num{14000}$\,K to \num{26000}\,K and $\log(g)=5$ or $6$, we were able to get a best fit using $\Teff=\num{26000}$\,K and $\logg=6.0$, as shown in Fig.~\ref{fig:spectra}.  
Note that the best fit values are at the extreme grid points of our sdBs models.
In particular, this model has a logarithmic helium abundance of $[\text{He}/\text{H}] = -1$, the lowest value in our grid, and many of the He~\textsc{I} lines present in the model are not seen or are weaker than predicted.
Thomas Kupfer (private communication), however, found $\Teff = \num{28400}$ $\pm 300$~K and $\log(g) = 5.57 \pm 0.05$ with a helium abundance of $-2.5 \pm 0.3$ using a more extensive range of sdB models.
The fitted radial velocities turn out to span a much smaller range ($11\pm 5\,\kms$ to $42\pm5\,\kms$) compared to PTF1 J2235+1428.

\section{Assessment of the Selection Procedure}
\label{sec:selection}
\subsection{Selection effects}\label{subsection:biases}
Some of the selection criteria used to identify periodic sources are not necessarily strict as they rely on manual inspection and filtering.
As a result, it is difficult to quantify the selection effects in our sample and we resort to exploiting qualitative clues on physical grounds.

Radial velocity searches favour binaries with a high primary-to-companion mass ratio, whose companion semi-amplitude goes as
\begin{equation}
K_2 \sim m_1/[P (m_1+m_2)^2]^{1/3},
\end{equation}
where $P$ is the orbital period, and $m_1$ and $m_2$ are the masses of the primary (WD) and companion, respectively \citep{hek}. 
On the other hand, this work is based on optical variability originated primarily from ellipsoidal variation and reflection. 
For both mechanisms, a high modulation amplitude strongly favours close binaries, as discussed in Section~\ref{subsection:period_determination}. 
Specifically, in contrast to radial velocity searches, they require one of the components (primarily for ellipsoidal modulation and secondarily for reflection) to be very close to filling its Roche lobe, with additional preference for a high companion-to-primary mass ratio for ellipsoidal modulation as shown in Eq.~\eqref{eq:parameters}.

To investigate the selection effects in a more quantitative fashion, we cross-checked our sample of periodic sources with 36 of the 58 PCEBs from \citet{go}, henceforth denoted as NGM, that have more than 20 PTF measurements.
It appears that the intersection between the two samples is null save one system, PTF1 J1559+0356.
NGM reported a period estimate of $2.266$~h, whereas we obtained $2.264$~h from L-S periodogram. 
The two estimates differ by one part per 500.
A closer look reveals that the NGM sources have fewer photometric measurements in PTF compared to our sample, with a median of 71 versus 153.
In addition, they are relatively fainter in PTF with a median magnitude of 18.4 versus 16.7, resulting in larger photometric errors.  
The combined effect made them more difficult to detect in a L-S periodogram, as evidenced by a median peak PSD of 8.0 versus 47.0.
Finally, NGM identified 65 of the 79 PCEBs via radial velocity measurements instead of photometry, and we expect the detection biases in the foregoing discussion to also contribute to the limited overlap between the two samples. 

\begin{figure}
\centering
\includegraphics[width=\columnwidth]{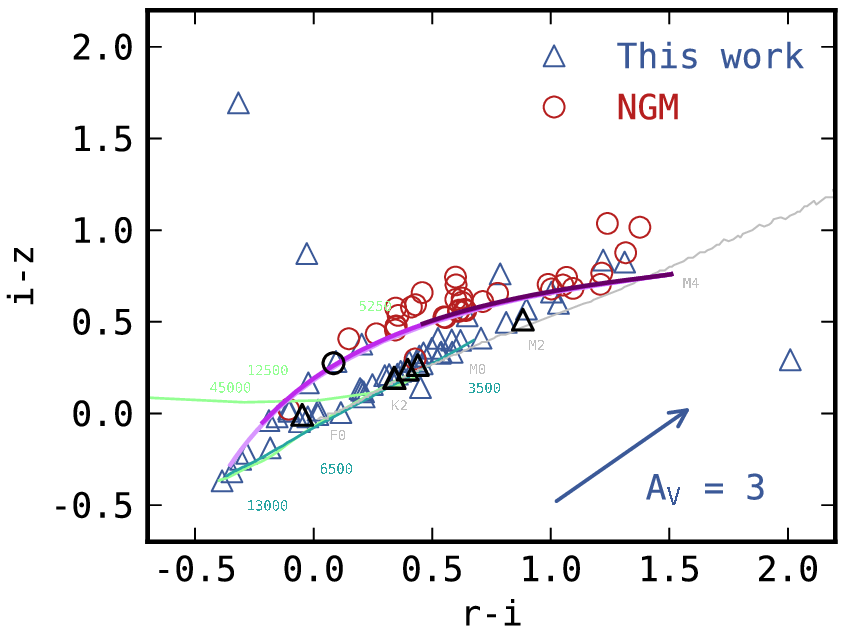}
\caption{Colour-colour plane for periodic variables in this work and in \citet{go} (NGM), displayed alongside theoretical WD tracks with $\log g = 8.0$ for He WDs (light green: DA; turquoise: DB) from \citet{ho} as well as MS stellar locus (gray) from \citet{co}.
A selected number of $T_{\text{eff}}$ values and MK spectral type labels are shown along the model tracks.
In addition, synthetic binaries comprising an M4 secondary and a DA primary with a mass of $0.4$, $0.7$, and $1.0M\odot$ are shown in shades of purple, from light to dark. 
Variables with $\log(\text{P}) > 0.5$~d are highlighted by bigger markers with black edges, and a reddening vector with $3$ magnitude of extinction was computed using extinction coefficients for SDSS filters from \citet{gir}.
}
 \label{fig:pc}
\end{figure}

In Table~\ref{table:p}, the majority of the periodic variables are in \citet{bi} only, and the lack of classification spectra gives rise to the possibility that the sample may contain sdO/sdB binaries instead of purely WD binaries. 
In particular, about $50\%$ of sdBs reside in close binaries, with either a WD or MS companion \citep{he}.
In general, subdwarfs have a larger radius compared to WDs and are typically more luminous, resulting in more pronounced reflection effect. 
To gain insights into the spectral type of the companion star, we show in Fig.~\ref{fig:pc} the SDSS $r-i$ and $i-z$ colours for the periodic variables in our sample, along with the DA and DB model tracks from \citet{ho} and the MS stellar locus from \citet{co}. 
For the sake of comparison, we also include the 36 NGM objects, which have larger $r-i$ and $i-z$ colours than our sample.
The discrepancy is expected since they are WD+MS binaries, which should result in a bias towards redder companions.
Quantitatively, we created synthetic WD+MS photometry by combining fluxes of a model M4 star and DAs of selected masses, all taken from the aforementioned model tracks.
The choice of binary components reflects the dominant spectral type in the NGM sample, as 14 of the 36 sources have been identified as DA+M4 systems \citep{go}.
As shown in Fig.~\ref{fig:pc}, the resulting binary tracks are in good agreement with the NGM sources, confirming our intuition. 
In addition, it is possible that our sample appears hotter on average, since a significant portion consists of variables potentially exhibiting reflection effect.

In summary, our sample complements those reported by NGM and \citet{ref:rebassa-mansergas+12} but cannot be considered complete due to selection effects.
However, future quantitative cross-comparison of the samples can serve as steps towards estimating selection effects in the hope of compiling a PCEB sample from which population characteristics, such as the intrinsic orbital period distribution, can be more accurately inferred.

\protect\footnotetext{http://www.astro.umontreal.ca/∼bergeron/CoolingModels/}

\subsection{Verification of selection fidelity}
\label{subsec:verify}
We investigated the false alarm probability of the present search by applying the same selection procedure on a sample of non-variable stars.
The SDSS Stripe 82 Standard Star Catalogue \citep{iv} contains photometrically identified standard stars located in stripe 82 ($20^\text{h} < \alpha_{\text{J2000.0}} < 4^\text{h}$ and $|\delta_{\text{J2000.0}}| < 1\overset{\circ}{.}266$).
Since SDSS has imaged this region over 10 times, the catalogue is sufficiently homogeneous \citep{iv}.
Additionally, since the median magnitude of these standard stars (19.61) is close to that of the sources from the three catalogues listed in Table~\ref{table:sources} (19.56), we arbitrarily selected 10000 stars from stripe 82 as our control sample. 
A query in the PTF data base returned 7764 objects with more than 20 data points, of which only 36 passed the machine filtering criteria (metrics: L-S PSD and amplitude).
By visually inspecting these 36 objects, we concluded that only one (PTF1 J2034--0021) is potentially outbursting.
Even though this object is faint in PTF with a median magnitude of $20.126$ and has only one point in outburst, we included it as a potential false positive.
We thus estimated the worst-case false alarm probability as $36/7764 \approx 0.5\%$.

Furthermore, we computed the probability that a non-periodic source was identified as periodic through a Monte Carlo simulation.
For each periodic candidate in our sample, we calculated the maximum PSD $\text{max(PSD)}_{\text{real}}$ using its PTF light curve. 
By shuffling the magnitudes while keeping each epoch, we produced a set of mock light curve and determined the corresponding peak PSD, denoted by $\text{max(PSD)}_{\text{mock}}$. 
A distribution of the mock L-S peaks was obtained by generating 1000 sets of such data.
By comparing $\text{max(PSD)}_{\text{real}}$ and each $\text{max(PSD)}_{\text{mock}}$, we constructed the null hypothesis probability ($H_0$: no periodicity) as $p = N/1000$, where $N$ is defined as the number of mock time series such that $\text{max(PSD)}_{\text{mock}} \ge \text{max(PSD)}_{\text{real}}$.
It turns out that we obtained $p = 0$ for all cases. 
One caveat is that the simulation would not detect a false period as identified by the L-S periodogram; it simply determines that the strongest peak in the periodogram is no chance occurrence.
Therefore, it is still possible that we have non-periodic signals in our sample due to artefacts such as diffraction spikes and harmonics of the sampling.
We have removed these to the best of our ability through a series of manual checks described in Section \ref{subsection:period_determination} and therefore believe that the vast majority of the reported periodic variables are genuine.

\section{Conclusions}
\label{sec:conclusion}
We have presented a sample of 59 periodic WD and/or sdB/sdO binary candidates from PTF light curves, of which {\newsrc} were not previously identified as periodic variables.
The sample also includes three newly identified eclipsing binaries.
Our work adds significantly to the roughly 100 previously identified PCEBs with orbital period estimates, providing additional individual sources for further characterization of component types, masses, and separations.
Knowledge of these binary parameters will aid future work towards estimating a bias-corrected population of PCEBs.  

The present sample was drawn from published catalogues of spectroscopically and photometrically selected WD candidates and relied on PTF data from higher galactic latitudes ($| b | \ge 20\,^{\circ}$).  
Since 2012, PTF has initiated an optical variability survey of the Galactic plane, and we expect future analysis of these data to yield an additional large sample of PCEB sources.

\section*{Acknowledgments}
Part of this work was performed by TAP while at the Aspen Center for Physics, which is supported by NSF Grant \#1066293.
WK and KBK were each funded by a Summer Undergraduate Research Fellowship at the California Institute of Technology. 

The CRTS survey is supported by the US National Science Foundation under grants AST-0909182.
Observations were obtained with the Samuel Oschin Telescope at the Palomar Observatory as part of the PTF project, a scientific collaboration between the California Institute of Technology, Columbia University, Las Cumbres Observatory, the Lawrence Berkeley National Laboratory, the National Energy Research Scientific Computing Center, the University of Oxford, and the Weizmann Institute of Science. 
Additional observations were obtained using the DBSP spectrograph on the Palomar 200-inch Hale telescope operated by Caltech.

We thank Nathaniel Butler for providing the main framework of the L-S \textsc{Python} script and Detlev Koester for providing us with his WD model atmospheres.   
We are in debt to Mike Yang for assisting in the retrieval of CRTS data and Ashish Mahabal for the verification of eclipses in CRTS light curves. 
Furthermore, the constructive comments from Eric Bellm, Thomas Kupfer, and Avi Shporer are much appreciated.
We would also like to thank the referee for careful reading of the manuscript and providing many valuable suggestions that certainly helped improve the paper.

WK thanks the productive discussions with Marten van Kerkwijk, as well as the time of Hung Yu (Ben) Ling and Alexander Papanicolaou, who never failed to provide seemingly useless advice that turned out as thought-provoking as caffeine. 
Last but not least, a special mention must go to Yi Cao, Adam Miller, and other members of the iPTF collaboration at Caltech, who generously shared their telescope time with us.

This research has made use of the SIMBAD data base, operated at CDS, Strasbourg, France, and has benefited much from NASA's Astrophysics Data System.

\appendix
\section{Folded Light Curves}\label{section:flc}
Here we present the PTF light curves for the 59 periodic WD candidates folded by the period given in Table~\ref{table:p}.
For systems with more than 200 measurements, a binned light curve (in gray) is also shown, where the width is scaled by the standard error of the points in each bin.

\begin{figure*}
 \centering
 \includegraphics[width=\textwidth]{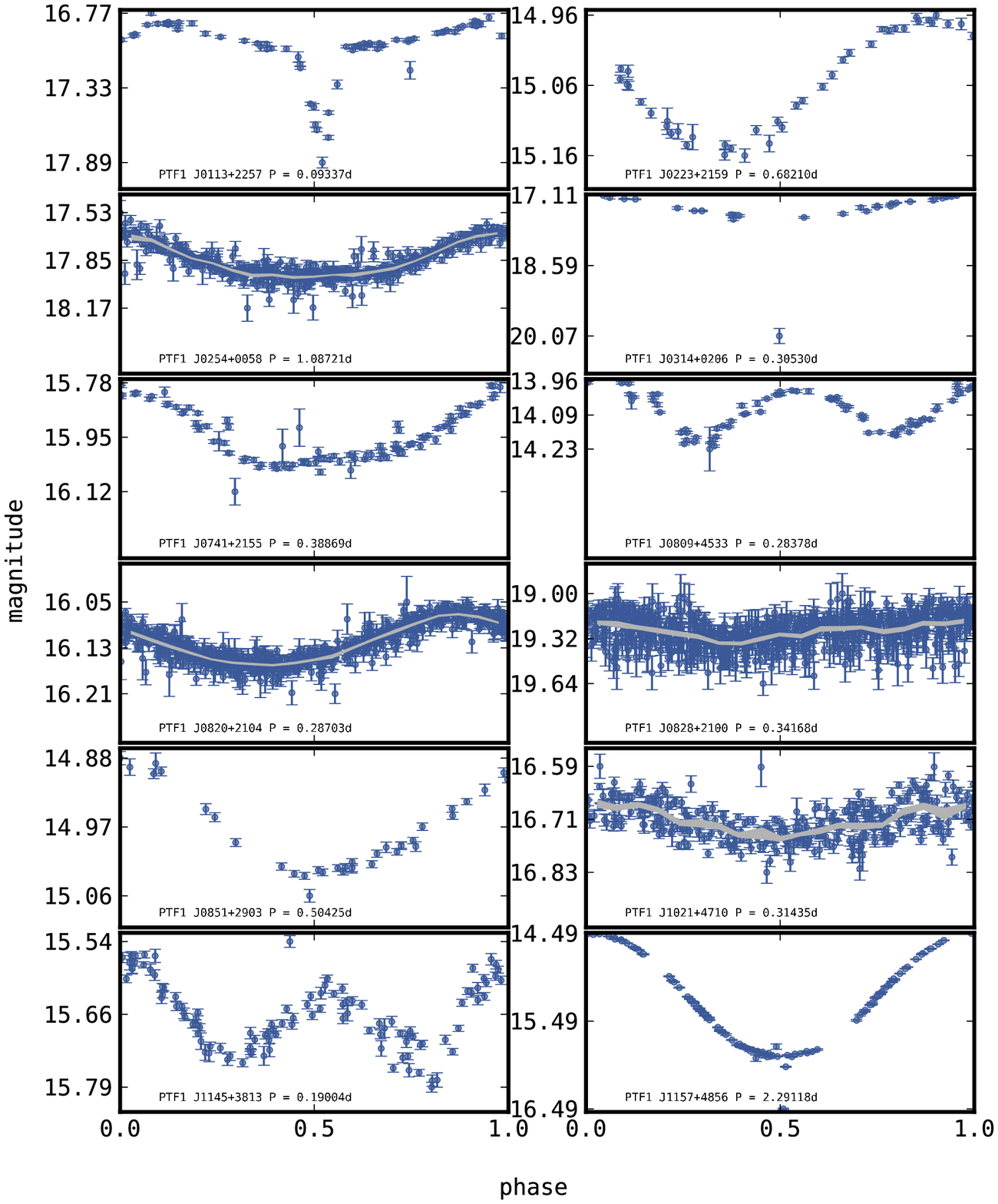}
 \caption{The 23 periodic variables where the difference of the PTF and CRTS best-fitting L-S periods is less than or on the order of $0.1\%$. 
 }
 \label{fig:a1}
\end{figure*}
\begin{figure*}
 \centering
 \includegraphics[width=\textwidth]{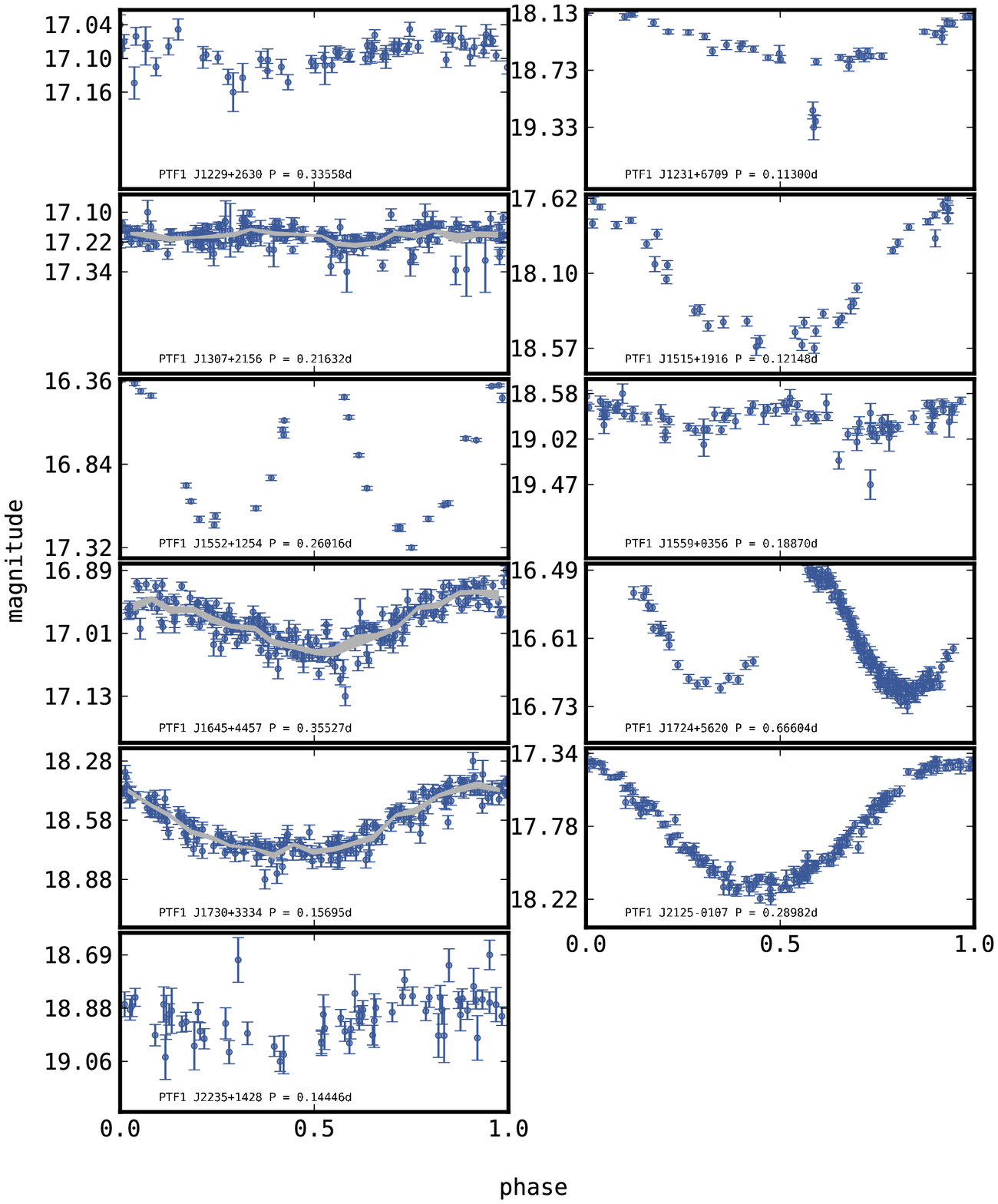}
 \caption{The 23 periodic variables - cont.}
 \label{fig:a2}
\end{figure*}
\begin{figure*}
 \centering
 \includegraphics[width=\textwidth]{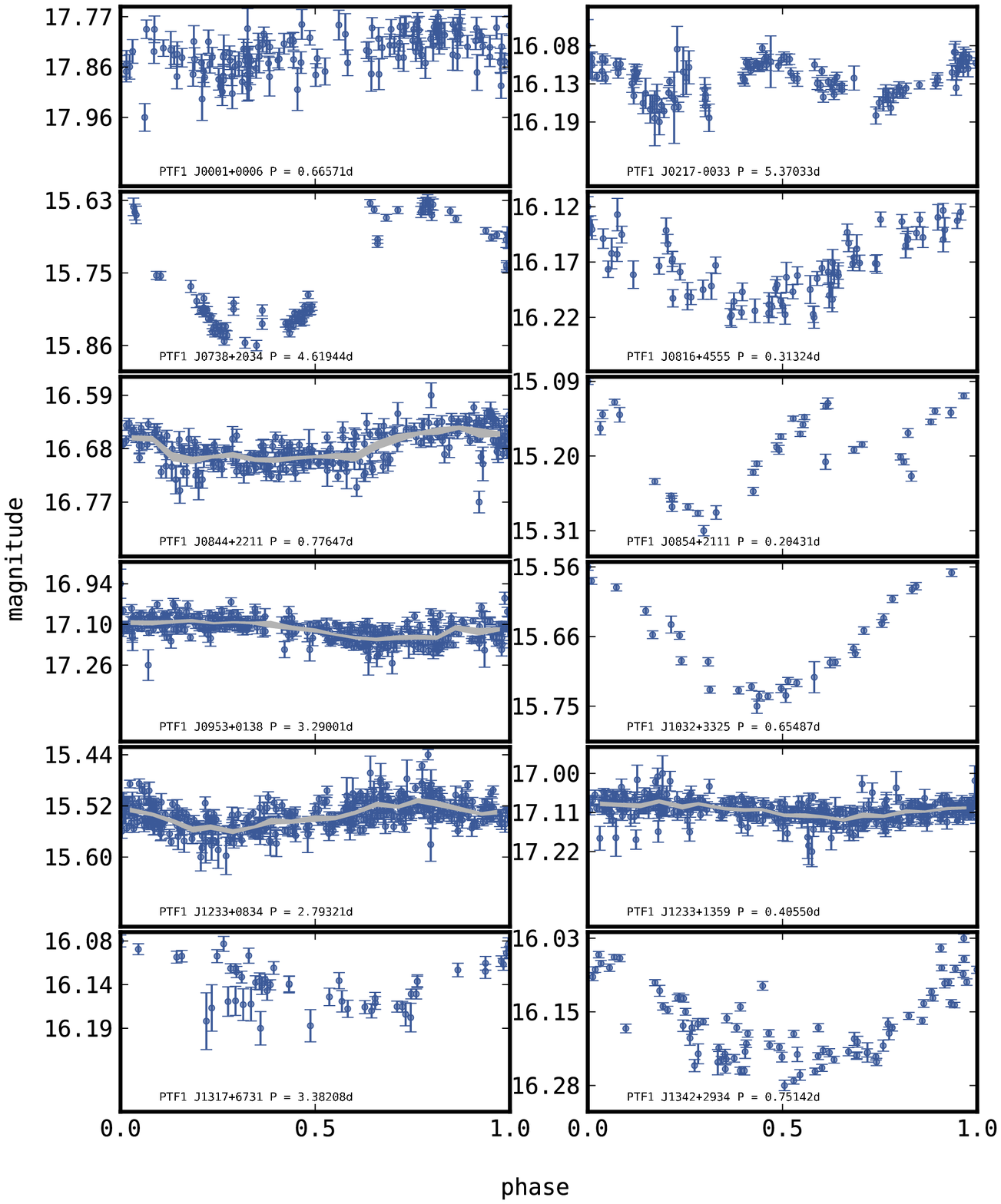}
 \caption{The remaining 36 periodic variables - I.}
 \label{fig:a3}
\end{figure*}
\begin{figure*}
 \centering
 \includegraphics[width=\textwidth]{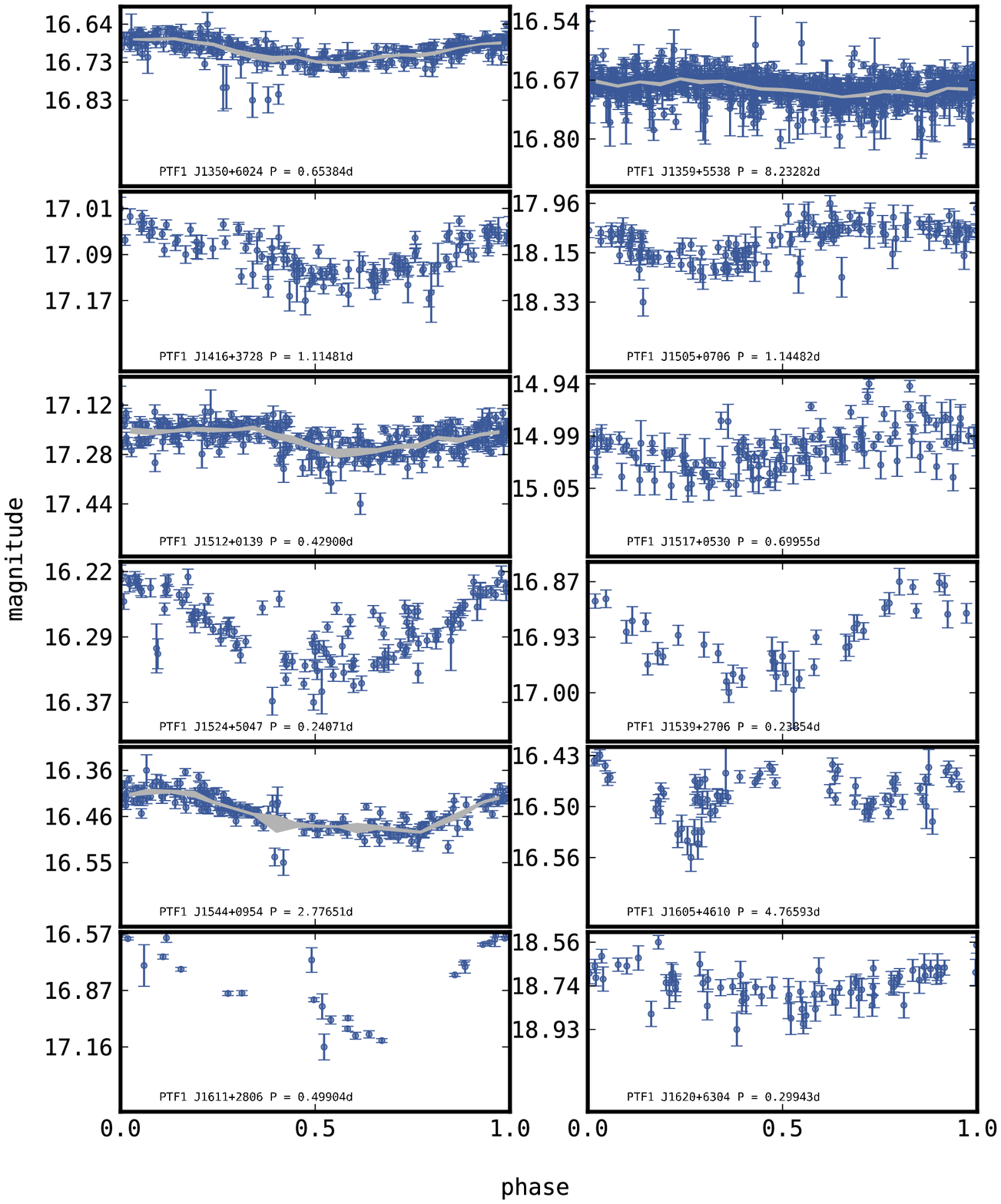}
 \caption{The remaining 36 periodic variables - II.}
 \label{fig:a4}
\end{figure*}
\begin{figure*}
 \centering
 \includegraphics[width=\textwidth]{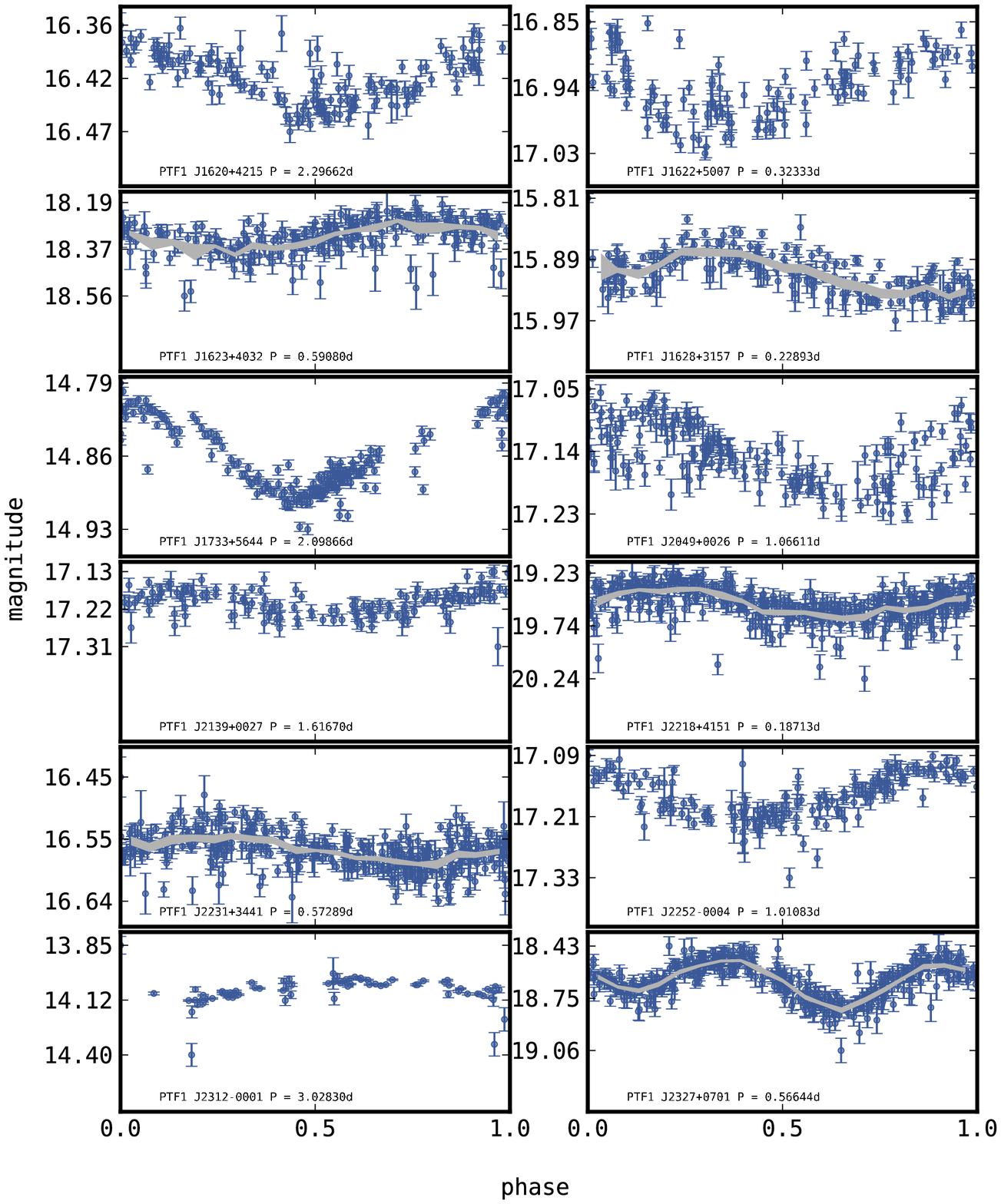}
 \caption{The remaining 36 periodic variables - III.}
 \label{fig:a5}
\end{figure*}

\end{document}